\documentclass[12pt]{iopart}
\usepackage{graphicx}
\usepackage{color}

\begin{document}

\title[Designing optimal feedback engines]{Designing optimal discrete-feedback thermodynamic engines}

\author{Jordan M. Horowitz and Juan M.~R.~Parrondo}

\address{Departamento de F\'isica At\'omica, Molecular y Nuclear and GISC, Universidad Complutense de Madrid, 28040 Madrid, Spain, EU}

\ead{parrondo@fis.ucm.es}

\date{\today}
\begin{abstract}
Feedback can be utilized to convert information  into useful work,
making it an effective tool for increasing the performance of
thermodynamic engines. Using feedback reversibility as a guiding principle, we
devise a method for designing optimal feedback protocols for
thermodynamic engines that extract all the information gained during
feedback as work.
Our method is based on the observation that in a feedback-reversible process the measurement and the time-reversal of the ensuing protocol both prepare the system in the same probabilistic state.
We illustrate the utility of our method with
two examples of  the multi-particle Szilard engine.
\end{abstract}


\maketitle

\section{Introduction}
An important application of feedback is to increase the performance of thermodynamic engines by converting the information gathered during feedback into mechanical work~\cite{Leff,Allahverdyan2008,Suzuki2009,Toyabe2010,Kim2011,Abreu2011,Vaikuntanathan2011}.
However, for feedback implemented discretely -- through a series of
feedback loops initiated at predetermined times -- the second law of
thermodynamics for discrete feedback limits the maximum amount of
work that can be extracted~\cite{Sagawa2008,Sagawa2010,Horowitz2010,Suzuki2010,Ponmurugan2010,Sagawa2011b,Sagawa2011}.
Namely, the average work extracted $\langle W\rangle$ during a
thermodynamic process with discrete feedback in which a system is
driven from one equilibrium state at temperature $T$ to another
equilibrium state at the same temperature is bounded by the
difference between the information gained during feedback $\langle
I\rangle$ and the average free energy difference $\langle\Delta
F\rangle$:
\begin{equation}\label{eq:GenSecLaw}
\langle W\rangle \le kT\langle I\rangle -\langle \Delta F\rangle,
\end{equation}
where $k$ is Boltzmann's constant. Here, $\langle I \rangle$ is the
mutual information  between the microscopic state of the system and
the measurement outcomes, and $\langle \Delta F\rangle$ is the
average free energy difference between the initial equilibrium state
and the final equilibrium state, which may differ for each
measurement outcome.
Notice  \eref{eq:GenSecLaw} is expressed in terms of the extracted work, since we have in mind applications to thermodynamic engines.
This differs from the more common convention of using the work done on the system, which is minus the work extracted~\cite{Sagawa2008,Sagawa2010,Horowitz2010,Suzuki2010,Ponmurugan2010,Sagawa2011b,Sagawa2011}.

\emph{Optimal} thermodynamic engines extract the maximum amount of
work, saturating the bound in \eref{eq:GenSecLaw} [$\langle
W\rangle =kT\langle I\rangle -\langle\Delta F\rangle$]. Their design
often proceeds in two steps. One first selects a physical observable
$M$  to be  measured. Then, associated to each measurement outcome
$m$,  one chooses a unique protocol for varying a set of external
parameters $\lambda$ during a time interval from $t=0$ to $\tau$, $\Lambda^m=\{\lambda_t^m\}_{t=0}^\tau$.
For the process to be optimal the collection of protocols $\{\Lambda^m\}$ must be
designed to extract as work all the information gained from the
measurement.

While at first it may not be obvious how to design a collection of optimal  protocols~\cite{Kim2011,Abreu2011}, there is a generic procedure for constructing such a collection given a physical observable $M$ \cite{Abreu2011,Jacobs2009,Erez2010,Hasegawa2010,Takara2010,Esposito2011}; specifically, the optimal protocol is to instantaneously switch the Hamiltonian immediately after the measurement -- through an instantaneous change of the external parameters -- so that the probabilistic state of the system conditioned on the measurement outcome is an equilibrium Boltzmann distribution with respect to the new Hamiltonian.
The external parameters are then reversibly adjusted to their final value, completing the protocol.
While such a protocol can always be constructed theoretically, it may be difficult to realize experimentally: one may need access to an infinite number of external parameters in order to affect the instantaneous switching of the Hamiltonian~\cite{Esposito2011}.
Furthermore, there are optimal protocols that cannot be constructed by implementing this generic procedure.
Hence, it is worthwhile to develop alternative procedures for engineering collections of optimal protocols.

In a recent article, we characterized optimal feedback processes, demonstrating that they are \emph{feedback reversible} -- indistinguishable from their time-reversals~\cite{Horowitz2011}.
There we pointed to the possibility of exploiting feedback reversibility in the design of optimal thermodynamic engines.
In this article, we take the next step by explicitly formulating a recipe for engineering a collection of optimal feedback protocols for a given observable $M$ using feedback reversibility as a guiding principle.
We present our method in \sref{sec:prep}, generalizing the generic procedure outlined in the previous paragraph.
We then illustrate our method in \sref{sec:ex} with
two pedagogical models inspired by the multi-particle Szilard
engine recently introduced in~\cite{Kim2011}, and subsequently analyzed in~\cite{Kim2011b}: a classical
two-particle Szilard engine with hard-core interactions, and a classical
$N$-particle Szilard engine with short-ranged, repulsive
interactions. In each model, we design a different
collection of feedback protocols, demonstrating the utility and
versatility of our method.
Concluding remarks are offered in \sref{sec:conclusion} with a view towards potential applications of our method to quantum feedback.

\section{Measurement and preparation}\label{sec:prep}

In this section, we describe a general method for designing optimal feedback protocols.
Our analysis is based on a theoretical framework characterizing the thermodynamics of feedback formulated in~\cite{Sagawa2010,Horowitz2010,Suzuki2010, Ponmurugan2010,Sagawa2011b,Sagawa2011,Horowitz2011}.

Consider a classical system whose position in phase space at
time $t$ is $z_t$. 
The system, initially in equilibrium at temperature $T$, is driven by varying  a set of external control parameters
$\lambda$ initially at $\lambda_0$ from time $t=0$ to $\tau$ using feedback. At time $t=t_m$,
an observable $M$ is measured whose outcomes $m$ occur randomly with probability $P(m|z_{t_m})$ depending only on the state of the system at the time of measurement $z_{t_m}$. The protocol,
denoted as $\Lambda^m=\{\lambda_t^m\}_{t=0}^\tau$, depends on the measurement outcome after time $t_m$.
Thermal fluctuations cause the system to trace out a random
trajectory  through phase space $\gamma=\{z_t\}_{t=0}^\tau$. The
work extracted along this trajectory is $W[\gamma; \Lambda^m]$, and the reduction
in our uncertainty due to the measurement is~\cite{Sagawa2010,Horowitz2010,Horowitz2011}
\begin{equation}\label{eq:I2}
I[\gamma;\Lambda^m]=\ln\frac{P(m|z_{t_m})}{P(m)},
\end{equation}
where $P(m)$ is the probability of obtaining measurement outcome $m$.
For error-free measurements, which we consider in our illustrative examples below, the measurement outcome is uniquely determined by the state of the system at the time of measurement.
Consequently, $P(m|z_{t_m})$ is always either zero or one.
When $P(m|z_{t_m})=1$, \eref{eq:I2} reduces to 
\begin{equation}\label{eq:I}
I[\gamma;\Lambda^m]=-\ln P(m).
\end{equation}
When $P(m|z_{t_m})=0$, \eref{eq:I2} is divergent; however, this divergence occurs with zero probability, and therefore does not contribute to the average in \eref{eq:GenSecLaw}.
Finally, the change in free energy from the initial equilibrium state, $F(\lambda_0)$, to the final equilibrium state, $F(\lambda^m_\tau)$, denoted as $\Delta F[\Lambda^m]=F(\lambda^m_\tau)-F(\lambda_0)$, is realization dependent, since the final external parameter value at time $\tau$ depends on the measurement outcome $m$.

Associated to the feedback process is  a distinct thermodynamic
process called the reverse process~\cite{Horowitz2010, Sagawa2011b,Horowitz2011}.
 The reverse process begins by first randomly
selecting a protocol $\Lambda^m$ according to $P(m)$. 
The system is then prepared in an equilibrium state at temperature $T$ with external parameters set to $\lambda^m_\tau$. From time
$t=0$ to $\tau$, the system is driven by varying the external
parameters according to the time-reversed conjugate protocol $\tilde{\Lambda}^m=\{{\tilde \lambda}_t\}_{t=0}^\tau$, where $\tilde\lambda^m_t=\lambda^m_{\tau-t}$.
For every trajectory
$\gamma=\{z_t\}_{t=0}^\tau$ of the forward process there is a
time-reversed conjugate trajectory
$\tilde\gamma=\{\tilde{z}_t\}_{t=0}^\tau$, where
$\tilde{z}_t=z_{\tau-t}^*$ and $*$ denotes momentum reversal.

A feedback process that is indistinguishable  from its reverse
process is called \emph{feedback reversible} \cite{Horowitz2011}. A useful microscopic
expression for the present considerations is in terms of the phase
space densities along the feedback process and the
corresponding reverse process.
Namely, the phase space density of the feedback  process at time $t$
conditioned on executing protocol $ \Lambda^m$,
$\rho(z_t|\Lambda^m)$, is identical to the phase space density in
the reverse process at time $ \tau-t$ conditioned on executing
protocol $ \tilde\Lambda^m$,
$\tilde\rho(\tilde{z}_{\tau-t}| \tilde\Lambda^m)$:
\begin{equation}\label{eq:reversible}
\rho(z_t|\Lambda^m)=\tilde\rho(\tilde{z}_{\tau-t}| \tilde\Lambda^m).
\end{equation}
Additionally,
\begin{equation}\label{eq:WIequal}
W[\gamma,\Lambda^m]=kTI[\gamma,\Lambda^m]-\Delta F[\Lambda^m]
\end{equation}
for every realization~\cite{Horowitz2011}.
For cyclic ($\Delta F=0$) feedback-reversible processes, such as our illustrative examples, \eref{eq:WIequal} is simply $W[\gamma,\Lambda^m]=kTI[\gamma,\Lambda^m]$.

We now utilize \eref{eq:reversible} and \eref{eq:WIequal} to
develop a method for designing optimal feedback processes (or
equivalently feedback-reversible processes). Our method is based on
the observation that  \eref{eq:reversible} has a noteworthy interpretation at the measurement time $t=t_m$:
\begin{equation}\label{eq:revMeas}
\rho(z_{t_m}|\Lambda^m)=\tilde\rho(\tilde{z}_{\tau-t_m}|\tilde\Lambda^m).
\end{equation}
Specifically, $\rho(z_{t_m}|\Lambda^m)$ is the phase space density of  the
system at the time of the measurement conditioned on implementing
protocol $\Lambda^m$; it represents our knowledge about the
microscopic state of the system immediately after the measurement.
We therefore refer to it as the \emph{post-measurement} state. The
right hand side of \eref{eq:revMeas},
$\tilde\rho(\tilde{z}_{\tau-t_m}| \tilde\Lambda^m)$, is the phase
space density at time $t=\tau-t_m$ produced by the reverse process
when protocol $ \tilde\Lambda^m$ 
 is executed;
it is the probabilistic state of the system prepared (or produced)
by using protocol $\tilde\Lambda^m$ in the reverse process. Thus,
we refer to $\tilde\rho(\tilde{z}_{\tau-t_m}|\tilde\Lambda^m)$ as the
\emph{prepared} state. 
With this terminology, \eref{eq:revMeas} states that for a process to be feedback reversible the state prepared
by the reverse process must be identical to the post-measurement
state. 
This insight is our main tool for
designing optimal feedback protocols. Instead of focusing on the
feedback process, we search for a protocol that prepares the
post-measurement state.
We call this procedure \emph{preparation}.
Once we have chosen our protocols, we can verify their effectiveness by checking the equality in \eref{eq:WIequal}; the deviation from equality in \eref{eq:WIequal} is a measure of the the reversibility of each of the protocols in $\{\Lambda^m\}$.

\section{Applications to the multi-particle Szilard engine}\label{sec:ex}

In this section, we apply the preparation method presented in \sref{sec:prep} to two classical extensions of the Szilard engine inspired by the quantum multi-particle Szilard engine considered by Kim \emph{et.~al.~}in \cite{Kim2011}.
In \sref{subsec:two}, we design a collection of optimal protocols for a classical Szilard engine composed of two square particles with hard-core interactions.
An $N$-particle Szilard engine consisting of ideal point particles with short-ranged, repulsive interactions is analyzed in \sref{subsec:many}.
In both examples, we verify that our protocols are optimal through analytic calculations of the work and information.

\subsection{Two-particle Szilard engine} \label{subsec:two}

\begin{figure}[htb]
\centering
\includegraphics[scale=0.35, angle=-90]{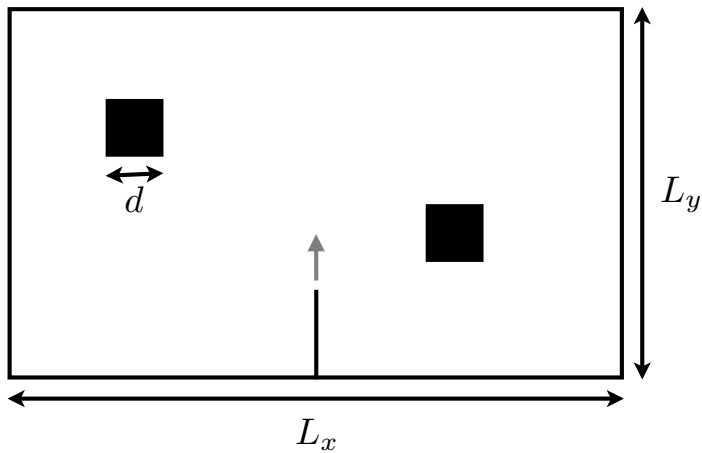}
\caption{Depiction of the two-particle Szilard engine composed of two square hard-core particles of width $d$ in a two-dimensional box of width $L_x$ and height $L_y$. Feedback protocols are initiated by infinitely slowly inserting a partition from the bottom edge of the box, as indicated by the vertical gray arrow, dividing the box into equal halves of width $L_x/2$.}
\label{fig:2Szilard}
\end{figure}

To illustrate the utility of our method, we now analyze a two-particle Szilard engine.
We have in mind two indistinguishable square hard-core
particles with linear dimension $d$ confined to a two-dimensional
box of width $L_x$ and height $L_y$, pictured in
\fref{fig:2Szilard}.
The particles have a hard-core interaction with the walls,  entailing
that the center of the particles must be at least a distance $d/2$
from the walls. The box is in weak thermal contact with a thermal
reservoir at temperature $kT=1$.

\begin{figure}[h!]
\centering
\includegraphics[scale=.3, angle=-90]{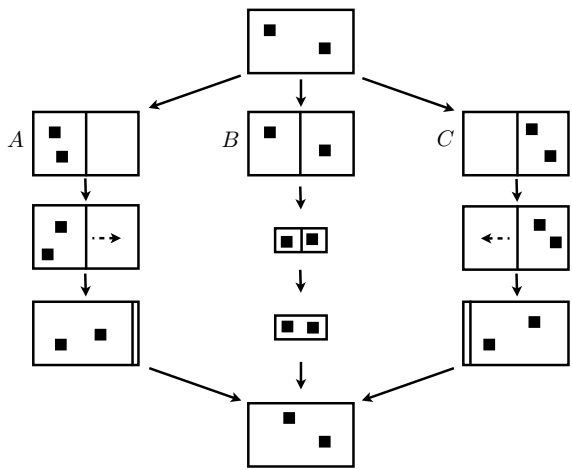}
\caption{Illustration of the three protocols executed in the two-particle Szilard engine associated to the three measurement outcomes $A$, $B$, and $C$.}
\label{fig:protocol}
\end{figure}

Work is extracted using a cyclic, isothermal feedback protocol
performed infinitely slowly, as illustrated in
\fref{fig:protocol}. Since the process is cyclic, $\langle\Delta
F\rangle=0$, and we only need to investigate the extracted work. In
addition, since the process is infinitely slow and isothermal, the
work can be expressed in terms of partition functions, as in~\cite{Parrondo2001}. 
There are two configurational partition
functions that will prove useful: the first, denoted $Z_2(x,y)$, is
the partition function for the state when both particles are in the
same box of width $x$ and height $y$; the second,
$\bar{Z}_2(x,y)$, is the partition function for the state where the
particles are in seperate boxes, each of width $x$ and height
$y$. The calculation of these partition functions is a
straightforward though lengthy exercise in integral calculus, which we outline in \ref{sec:appendix}.

We initiate the feedback protocol with the engine in thermal equilibrium at temperature $kT=1$. 
We then infinitely
slowly insert a thin partition from below,  dividing the box into
two equal halves along the horizontal direction, as depicted in
\fref{fig:2Szilard}.
Because the particles are hard-bodied and of finite size, the insertion of the partition extracts work.
As we slowly insert the partition, the system remains in equilibrium and able to explore its entire phase space
until the leading tip of the partition is one particle length $d$ from the box's top wall.
At which point,  the particles are too large to pass between the left and right half of the box.
At that moment, each particle becomes trapped in one half of the box;
either they both become trapped in the same half of the box, or each is trapped in a separate half of the box.
The partition function at that moment, being a sum over all distinct microscopic configurations, is then the sum of the partition function when they both become trapped in the left (or right) half, $Z_2(L_x/2,L_y)$, plus the partition function when they become trapped in separate halves, $\bar{Z}_2(L_x/2,L_y)$: $2Z_2(L_x/2,L_y)+\bar{Z}_2(L_x/2,L_y)$.
The work extracted up to  that instant is determined from the ratio of the partition function at that moment to the initial partition function $Z_2(L_x,L_y)$ as
\begin{equation}\label{eq:Win}
W_{\rm part}(L_x,L_y)=\ln\left[\frac{2Z_2(L_x/2,L_y)+\bar{Z}_2(L_x/2,L_y)}{Z_2(L_x,L_y)}\right].
\end{equation}
Once the distance between the leading tip of the partition and the far wall of the box is less than $d$, neither particle is able to fit in the space between the tip and the wall.
The partition's tip is no longer able to push on the particles, and as a result no additional work beyond that in \eref{eq:Win} is extracted.

Next, we measure in which half of the box the two particles  are
located. There are three outcomes, which we label $A$, $B$, and $C$,
see \fref{fig:protocol}. Outcomes $A$ and $C$ occur when both particles are found in the same half of the
box, whereas outcome $B$ occurs when each particle is found in a
separate half of the box. Since the partition functions $Z_2$ and
$\bar{Z}_2$ count the number of distinct microscopic configurations,
we can express the change in uncertainties associated to each
outcome by inserting these partition functions into
\eref{eq:I}:
\begin{eqnarray}
\label{eq:IA}
I_A=I_C=-\ln\left[\frac{Z_2(L_x/2,L_y)}{2Z_2(L_x/2,L_y)+\bar{Z}_2(L_x/2,L_y)}\right], \\
\label{eq:IB}
I_B=-\ln\left[\frac{\bar{Z}_2(L_x/2,L_y)}{2Z_2(L_x/2,L_y)+\bar{Z}_2(L_x/2,L_y)}\right].
\end{eqnarray}

If both particles are found in the same half of the box (outcome $A$ or $C$),  
the optimal protocol is to quasi-statically shift the partition to the
opposite end of the box, as in the single-particle Szilard engine \cite{Szilard1964}, extracting work
\begin{equation}\label{eq:WorkExp}
W_{\rm shift}=\ln\left[\frac{Z_2(L_x,L_y)}{Z_2(L_x/2,L_y)}\right].
\end{equation}
Summing \eref{eq:Win} and \eref{eq:WorkExp}, we find that the work extracted during the feedback protocol associated to measurement outcome $A$ (or $C$) is
\begin{eqnarray}
W_A&=W_{\rm part}(L_x,L_y)+W_{\rm shift} \\
&=\ln\left[\frac{2Z_2(L_x/2,L_y)+\bar{Z}_2(L_x/2,L_y)}{Z_2(L_x/2,L_y)}\right],
\end{eqnarray}
which equals $I_A$ in \eref{eq:IA}. 
Thus, according to \eref{eq:WIequal} this protocol is optimal as expected, since this protocol when run in reverse clearly prepares the post-measurement state conditioned on $A$.

When each particle is found in a separate half of  the box (outcome
$B$), the optimal protocol is less clear. 
The motion of the piston in either direction requires work rather than extracts it.
Kim \emph{et.~al.,} for instance, opt to extract the partition without obtaining any useful work~\cite{Kim2011}:
the information in the measurement is wasted.
However, our discussion in \sref{sec:prep} suggests a way to design an optimal cyclic protocol:
the protocol must drive the system from the state post measurement outcome $B$ back to the initial state and when run in reverse must prepare the state associated to outcome $B$ by segregating each particle into a different half of the box.
When the particles do not interact, there  is no obvious
optimal protocol.
However, in our model we can exploit the particle interactions.
Specifically due to the hard-core interactions, there is a greater likelihood of trapping the particles in separate halves of
the box upon inserting the partition when the box is smaller.
This observation suggests the following protocol executed in response to measurement outcome $B$.

\begin{figure}[htb]
\centering
\includegraphics[scale=.35, angle = 90]{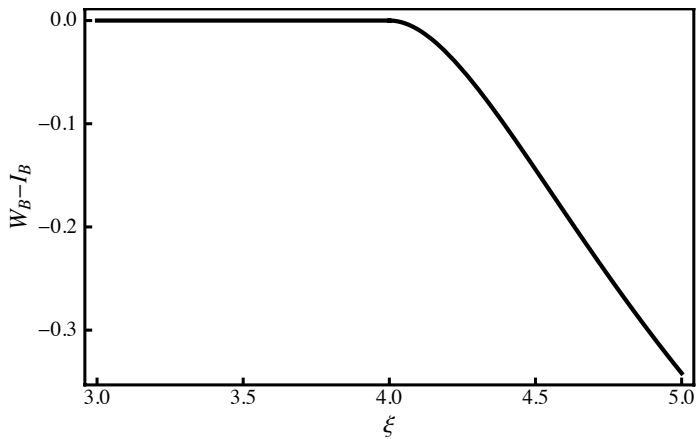}
\caption{Plot of the deviation from reversibility $W_B-I_B$ for the two-particle Szilard engine protocol implemented in response to measuring each particle in a separate half of the box (outcome $B$) as a function of the box size parameter $\xi=l_x/d=2l_y/d$.}
\label{fig:prep}
\end{figure}

After the partition is inserted, we
infinitely slowly compress the box until its width is $l_x>2d$ and
its height is $l_y>d$. The extracted work during compression is
\begin{equation}\label{eq:Wcomp}
W_{\rm comp}=\ln\left[\frac{\bar{Z}_2(l_x/2,l_y)}{\bar{Z}_2(L_x/2,L_y)}\right].
\end{equation}
Next, the partition is removed infinitely slowly, extracting $-W_{\rm part}(l_x,l_y)$ [see \eref{eq:Win}] work.
Finally, the box is expanded back to its original size extracting
\begin{equation}\label{eq:Wexp2}
W_{\rm exp}=\ln\left[\frac{Z_2(L_x,L_y)}{Z_2(l_x,l_y)}\right].
\end{equation}
Combining the sum of \eref{eq:Win}, \eref{eq:Wcomp}, \eref{eq:Wexp2}, and $-W_{\rm part}(l_x,l_y)$, with \eref{eq:IB}, we find, after a simple algebraic manipulation, that the deviation from reversibility [cf.~\eref{eq:WIequal}] can be expressed as
\begin{equation}\label{eq:WminusI}
W_B-I_B=-\ln\left[1+2\frac{Z_2(l_x/2,l_y)}{\bar{Z}_2(l_x/2,l_y)}\right].
\end{equation}
Note that $W_B-I_B$ only depends on the size of the compressed box with dimensions $l_x\times l_y$.
To investigate the reversibility of our protocol, we study the dependence of $W_B-I_B$ on the compressed box size.
To simplify our analysis, we only consider boxes such that $l_x=2l_y$.
In  \fref{fig:prep}, we plot  $W_B-I_B$ as a function of the box size parameter  $\xi=l_x/d=2l_y/d$.
The smaller $\xi$ the smaller the box. Notice that $W_B-I_B<0$. 
We also observe that the process becomes reversible ($W_B-I_B=0$)  when $\xi<4$ ($l_x<4d$
and $l_y<2d$);  the box is so small when $\xi<4$ that both
particles cannot fit into the same half of the box. Consequently,
when the partition is inserted during the reverse process each
particle is confined to a separate half of the box, preparing
the post-measurement state with probability one.

\begin{figure}[htb]
\centering
\includegraphics[scale=.35, angle = -90]{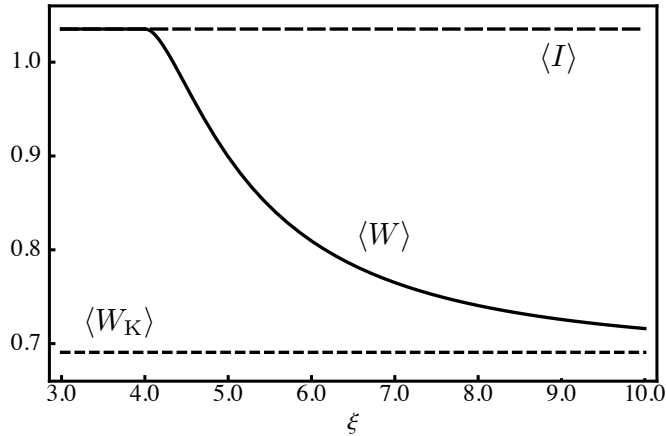}
\caption{ Plot comparing the total average work extracted $\langle W\rangle$ (solid) in our two-particle Szilard engine protocol to the information $\langle I\rangle$ (long dashed) as a function of the box size parameter $\xi=l_x/d=2l_y/d$ for a box initially of size  $L_x=20d$ by $L_y=10d$.
For reference, we have included the total average work that would have been extracted using the protocol introduced by Kim~{\it et al.}~in Ref.~\cite{Kim2011}, $\langle W_{\rm K}\rangle$ (dashed).}
\label{fig:totWork}
\end{figure}

 To confirm that our protocol can be optimal, we plot in \fref{fig:totWork} the total average work extracted $\langle W\rangle = P_AW_A+P_BW_B+P_CW_C$ --  where $P_j$ is the probability to implement protocol $j=A,B,C$ -- as a function of the box size parameter $\xi$.
 Again, we see that when $\xi<4$ our protocol becomes optimal: $\langle W\rangle = \langle I\rangle$.
For comparison, we have included in \fref{fig:totWork} the work extracted when implementing the protocol proposed in Ref.~\cite{Kim2011}, $\langle W_{\rm K}\rangle$, where the partition is slowly removed in response to outcome $B$.

Further insight can be gained  by noting that the ratio $Z_2/\bar{Z}_2$ in \eref{eq:WminusI}, which controls the degree of reversibility, has a simple physical interpretation in terms of the change in free energy during an irreversible mixing of two indistinguishable particles, each in separate boxes of sizes $l_x/2\times l_y$, into one box of the same size, $l_x/2\times l_y$:
\begin{equation}
\Delta F_{\rm mix}=-\ln\left[\frac{Z_2(l_x/2,l_y)}{\bar{Z}_2(l_x/2,l_y)}\right].
\end{equation}
Thus, this protocol is reversible when there is an infinite free energy
difference between the states in which both particles are in the same
box and where each particle is in a separate box.
For an ideal gas $\Delta F_{\rm mix}=\ln2$: two indistinguishable
ideal gas particles confined to the same box have half as many
distinct microscopic configurations than when they are in seperate
boxes.
For ideal gases our protocol is not optimal ($\Delta F_{\rm mix}\neq\infty$ and $W_B-I_B\neq0$),
as it exploits particle interactions.
Nevertheless, there may exist other protocols that are optimal for ideal gases.
In particular, such a collection could be devised using the generic procedure outlined in the Introduction, where the Hamiltonain is instantaneously switched immediately after the measurement so that the post-measurement state is described by an equilibrium Boltzmann distribution with respect to the new Hamiltonian~\cite{Abreu2011,Jacobs2009,Erez2010,Hasegawa2010,Takara2010,Esposito2011}; however, this new Hamiltonian would contain an interaction potential that forces the particles to segregate themselves into opposite halves of the box.

\subsection{$N$-particle Szilard engine}\label{subsec:many}

As a final illustration, we present  an optimal feedback  protocol
for a classical $N$-particle Szilard engine. Consider $N$ indistinguishable, classical,
point particles with short-ranged, repulsive interactions confined
to a box of volume $V$ in weak thermal contact with a thermal
reservoir at temperature $kT=1$. The protocol begins by  quickly and isothermally
inserting an infinitely thin partition into the box dividing it into
two equal halves of volume $V/2$. 
Since this is performed rapidly and the particles are infinitely small, the particles never have an opportunity to interact with the partition implying that this insertion requires no work.
We then
measure the number of particles in the left half of the box. Based
on the outcome, we implement a cyclic, isothermal feedback protocol.

\begin{figure}[htb]
\centering
\includegraphics[scale=0.37,angle=-90]{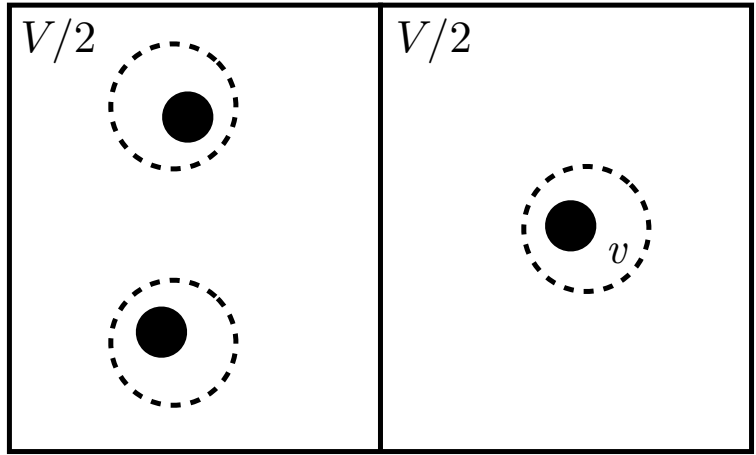}
\caption{Depiction of the $3$-particle Szilard engine protocol in a box of volume $V$ after having turned on the trapping potentials when $2$ particles were found in the left half the box.
Each of the three particles, pictured as black circles, is confined to a distinct trapping potential of volume $\emph{v}$, pictured as a dashed circle.}
\label{fig:Nparticle}
\end{figure}

The change in uncertainty when $n$ particles are found in the  left
half of the box ($N-n$ particles in the right half) is, from~\eref{eq:I},
\begin{equation}\label{eq:In}
I_n=-\ln\left[\frac{1}{2^N}\frac{N!}{n!(N-n)!}\right].
\end{equation}
This information can be extracted completely as work by implementing
the following protocol. 
First, we slowly lower $n$ ($N-n$) localized
potential minima or  trapping potentials to a  depth $E$ in the left  (right) half of
the box. The trapping potentials are assumed to be deep compared to
the thermal energy ($E\gg kT$), but shallow compared to the interaction energy;
so that only one particle is confined in each trapping potential, as
depicted in \fref{fig:Nparticle}.
The partition is then quickly removed, and the trapping potentials are slowly turned off.

Work is only extracted when the trapping potentials are turned on or
off. Since these processes are very slow, the work extracted can be
computed in terms of partition functions. Assuming that the
volume $V$ of the box is large compared with the interaction length,
we can approximate the configurational partition function for the
equilbrium state prior to inserting the partition as
\begin{equation}
Z(V)=\frac{V^N}{N!}.
\end{equation} 
After making the measurement and finding $n$ particles in the left half of the box, the configurational partition function is
\begin{equation}
Z_n(V)=\frac{1}{n!(N-n)!}\left(\frac{V}{2}\right)^N.
\end{equation}
After lowering the trapping potentials to a depth $E$ each particle is confined to a unique trapping potential of volume $\emph{v}$.
At which point the configurational partition function  is
\begin{equation}
\bar{Z}_n(\emph{v})=\emph{v}^Ne^{-NE}.
\end{equation}
In terms of these partion functions, the work extracted while trapping the particles is
\begin{equation}\label{eq:trap}
W_{\rm trap}=\ln\left[\frac{\bar{Z}_n(\emph{v})}{Z_n(V)}\right]=\ln\left[2^N\left(\frac{\emph{v}}{V}\right)^Nn!(N-n)!e^{-NE}\right],
\end{equation}
and the work extracted when the trapping potentials are turned off is
\begin{equation}\label{eq:off}
W_{\rm off}=\ln\left[\frac{Z(V)}{\bar{Z}_n(\emph{v})}\right]=\ln\left[\frac{1}{N!}\left(\frac{V}{\emph{v}}\right)^Ne^{NE}\right].
\end{equation}
Summing \eref{eq:trap} and \eref{eq:off}, we find the total work to be
\begin{equation}
W_n=W_{\rm trap}+W_{\rm off}=\ln\left[2^N\frac{n!(N-n)!}{N!}\right],
\end{equation}
which is independent of $E$ and is equal to the change in uncertainty $I_n$ in \eref{eq:In}.
This protocol is optimal and feedback reversible; run in reverse the protocol confines exactly $n$ particles in the left half with certainty.

At first it may be surprising that work can be extracted from  this
protocol, since we are mearly adding and then removing potential
minima. However, net work can be extracted, since the work extracted while slowly turning on or off a
trapping potential depends on the total volume accesible to the
particles. To see this, consider the simplest scenario of turning off
one trapping potential with one particle confined to a box of volume
$V$.
As the depth of the potential minimum becomes shallower, work  is
done on the particle until it escapes from the range of the trapping
potential. Once the particle leaves, turning off the potential requires no additional work until the particle returns.
The time for the particle to return depends on the size of the box. 
For a box of larger volume, the time to return is longer, and the process requires less work. 
Going back to the $N$-particle protocol, the work extracted while turning on the trapping potentials after
the partition has been inserted -- when the available volume for each
particle is $V/2$ -- is more than the work done during the final step as
the trapping potentials are removed, because the volume
$V$ available for the particles to explore is larger.

When the number of trapping potentials is not equal to the number of particles $N$, this protocol is no longer optimal.
The reason being that work can only be extracted when a particle can fall into a potential being lowered; the more trapping potentials a particle has access to, the more work that can be extracted.
If there were less trapping potentials then particles, overall less work would be extracted; as there would be fewer sites where energy was being removed.
If more than $N$ trapping potentials are lowered, we are able to extract additional work.
However, after the partition is removed, each particle can explore an even greater number of trapping potentials; the work to turn off the potentials would exceed that extracted by turning them on.

\section{Conclusion}\label{sec:conclusion}

Feedback-reversible processes are optimal, converting all the information acquired through feedback into work.
In this article, we formulated a strategy, called preparation, for designing a collection of optimal protocols given a measured physical observable.
In the preparation method, optimal protocols are selected by searching for an external parameter protocol whose time-reversal prepares the post-measurement state.
To highlight the utility of the preparation method, we applied it to two pedagogical examples -- a two- and $N$-particle Szilard engine -- exhibiting a distinct collection of optimal protocols for each.
In both examples, we addressed the simplest scenario of error-free measurements.
When there are measurement errors -- for example, if in the $N$-particle Szilard engine (\sref{subsec:many}), there were a chance to miscount the number of particles in the left half of the box -- the preparation method still provides a useful procedure for selecting an optimal protocol. 
Furthermore, each of our optimal protocols contained at least one infinitely slow step.
This is unavoidable as the process must be reversible before and after any measurements.
Consequently, our method does not strictly apply to finite-time processes.
However, the preparation method may still provide insight into the design of optimal finite-time processes, since an optimal finite-time protocol, roughly speaking, is as close to reversible as possible~\cite{Abreu2011,Schmiedl2007}.

Generally, we expect the preparation method to be of use whenever the external parameter protocol forces a symmetry breaking in the system prior to the measurement, such as the insertion of the partition in the Szilard engine.
Consider a thermodynamic process ${\cal P}$ during which a system is driven from an initial equilibrium state $A$ through  a critical point, where the system chooses among several phases or macroscopic states $B_i$ with probability $p_i$. 
In addition, suppose there exists a collection of processes ${\cal P}'_i$ during which the symmetry is broken forcibly (not spontaneously), driving the system from $A$ to $B_i$ with probability one.
Then, according to our recipe this spontaneous symmetry breaking transition can be exploited using the following optimal feedback protocol: start in state $A$, execute process ${\cal P}$, measure which state $B_i$ resulted from the symmetry breaking, and then run the
corresponding process ${\cal P}_i^\prime$ in reverse to drive the system back to its initial
state $A$. 
By construction, this process prepares the post-measurement state with unit probability, and therefore extracts as work $\langle W\rangle= - kT\sum_i p_i\log p_i$, which is $kT$ times the information gained in the
measurement, $\langle I\rangle= - \sum_i p_i\log p_i$.
One interesting instance of this setup is the Ising model, where a measurement of the system's total magnetization after the symmetry breaking phase transition between the paramagnetic and ferromagnetic states can be exploited to extract work.
This information can be utilized by modifying an external  magnetic field, as demonstrated in~\cite{Parrondo2001}.

In the introduction, we outlined a general procedure for preparing a collection of optimal protocols, original presented in \cite{Abreu2011,Jacobs2009,Erez2010,Hasegawa2010,Takara2010,Esposito2011}, in which the Hamiltonian is instantaneously changed immediately following the measurement in order to make the post-measurment state an equilibrium Boltzmann distribution, followed by a reversible switching of the external parameters to their final values.
These protocols prepare the post-measurement states; as such this generic procedure is a special case of the preparation method developed here.
Though, the implementation of the preparation method can lead to a wider variety of protocols.
Take for example the two-particle Szilard engine discussed in \sref{subsec:two}.
Imagine we make a measurement and find outcome $B$, where each particle is confined to a separate half of the box.
Let $\rho_B(z)$ denote the phase space density conditioned on this measurement outcome.
In the generic procedure, immediately after the measurement we would change the Hamiltonian to $H_B(z)=-\ln\rho_B(z)$, which is a strange Hamiltonian that assigns infinite energy to configurations where both particles are in the same half of the box.
In contrast, the preparation method led to a physically realizable protocol, in which we vary the size of the box.

Finally, we formulated the preparation method only for classical systems.
Though, the second law of thermodynamics for discrete feedback was originally predicted for quantum evolutions~\cite{Sagawa2008}.
Its mathematical structure resembles the classical version, which suggests that feedback-reversible processes are also optimal quantum feedback protocols and that the preparation method would also apply to quantum feedback engines.
Applications of the preparation method to quantum systems holds interesting possibilities.
For example, in both the classical multi-particle Szilard engines analyzed here, the optimal protocols required repulsive particle interactions.
In a quantum multi-particle Szilard engine composed of fermions, the Pauli exclusion principle induces a repulsive interaction of purely quantum origin, which could be exploited to develop a collection of optimal feedback protocols.

\ack

We acknowledge Hal Tasaki for suggesting the $N$-particle Szilard engine protocol.
Financial support for this project came from Grant MOSAICO (Spanish Government) and MODELICO (Comunidad de Madrid).

\appendix

\section{Partition functions for two square hard-core particles in a two-dimensional box}\label{sec:appendix}

In this appendix, we report the configurational partition functions employed in Sect.~\ref{subsec:two} for a gas composed of two square particles of width $d$ with hard-core interactions confined to a two-dimensional box of width $L_x$ and height $L_y$.
The partition function for hard-core particles is the number of distinct microscopic configurations subject to the constraint that the centers of the particle be separated by a distance of at least $d$. 
In addition, the particles have a hard-core interaction with the walls enclosing the box, with the result that the center of each particle must be at least a distance $d/2$ from the edges of the box.

Two partition functions are utlized in our analysis in \sref{subsec:two}. 
The first is the partition function for the equilibrium state when each particle is confined to separate box of dimensions $L_x \times L_y$:
\begin{eqnarray}
\bar{Z}_2(L_x,L_y)&=\int_{d/2}^{L_x-d/2}dx_1\, \int_{d/2}^{L_y-d/2}dy_1\, \int_{d/2}^{L_x-d/2}dx_2\,  \int_{d/2}^{L_x-d/2}dy_2  \\
&=(L_x-d)^2(L_y-d)^2.
\end{eqnarray}
The second is for the equilibrium state when both particles are confined to the same box of dimensions $L_x \times L_y$. 
This partition function can be expressed as the integral
\begin{eqnarray}
\nonumber
\fl Z_2(L_x,L_y)=&\frac{1}{2}\int_{d/2}^{L_x-d/2}dx_1\, \int_{d/2}^{L_y-d/2}dy_1\, \int_{d/2}^{L_x-d/2}dx_2\, \int_{d/2}^{L_y-d/2}dy_2\, \\ \nonumber
&\times[\Theta(|x_1-x_2|-d)+\Theta(|y_1-y_2|-d)-\Theta(|x_1-x_2|-d)\Theta(|y_1-y_2|-d)],
\end{eqnarray}
where $\Theta(x)$ is the Heaviside step function and the preceding factor of $1/2$ is included because the particles are indistinguishable.
The calculation of the above integral can be performed using standard methods of integral calculus, with the result,  assuming $L_x>2d$,
\begin{equation}
\fl
Z_2(L_x,L_y)=
\left\{
\begin{array}{ll}
\frac{1}{2}(L_x-2d)^2(L_y-2d)^2+2d(L_x-2d)(L_y-2d)(L_x+L_y-4d) \\
\, \, \, +d^2\left[(L_x-2d)^2+(L_y-2d)^2\right], & L_y \ge 2d \\
\frac{1}{2}(L_y-d)^2(L_x-2d)^2, & d\le L_y< 2d
\end{array}
\right. .
\end{equation}

\bibliographystyle{iopart-num}
\bibliography{Feedback,PhysicsTexts}

\end{document}